\documentclass[11pt]{article}
\usepackage{a4wide}
\usepackage{amsmath} 
\usepackage{amssymb}

\newtheorem{theorem}{Theorem}[section]

\newtheorem{corollary}[theorem]{Corollary}
\newtheorem{definition}[theorem]{Definition}
\newtheorem{example}[theorem]{Example}
 
\newtheorem{lemma}[theorem]{Lemma}

\newenvironment{proof} {\noindent {\it Proof}.  } {\hfill $\diamond$}

\newcommand{\ds}{\displaystyle}

\newcommand{\ignore}[1]{}

\DeclareMathOperator{\depth}{depth}

\DeclareMathOperator{\ldepth}{ldepth}

\begin{document}

\title{Depth as Randomness Deficiency}

\author{Lu\'is Antunes\thanks{
Email: {\tt lfa@ncc.up.pt}.
Web: {\tt http://www.ncc.up.pt/\~{ }lfa}.
Adress: {\tt Departamento de Ci\^encia de Computadores. Rua Campo Alegre,
1021/1055, 4169 - 007 PORTO, PORTUGAL}
}\\
{\it Computer Science Department}\\
{\it University of Porto}
\and
Armando Matos\thanks{
Email: {\tt acm@ncc.up.pt}.
Web: {\tt http://www.ncc.up.pt/\~{ }acm}.}\\
{\it Computer Science Department}\\{
\it University of Porto}
\and
Andr\'e Souto\thanks{
Email: {\tt andresouto@dcc.fc.up.pt}.
Web: {\tt http://www.ncc.up.pt/\~{ }andresouto}.}\\
{\it Computer Science Department}\\
{\it University of Porto}\footnote{The authors from University of Porto are partially supported  by KCrypt (POSC/EIA/60819/2004) and funds  granted to LIACC  through   the  Programa  de Financiamento Plurianual, Funda\c{c}\~ao   para  a  Ci\^encia  e Tecnologia  and Programa POSI}\\
\and
Paul Vit\'anyi\thanks{
Email: {\tt Paul.Vitanyi@cwi.nl}.
Web: {\tt http://homepages.cwi.nl/\~{ }paulv/.}}\\
{\it CWI and}\\
{\it Computer Science Department}\\
{\it  University of Amsterdam}
}

\maketitle

\begin{abstract}
Depth of an object concerns a tradeoff between computation time and
excess of program length over the shortest program length required to
obtain the object. It gives an unconditional lower bound on the
computation time from a given program in absence of auxiliary
information. Variants known as logical depth and computational depth
are expressed in Kolmogorov complexity theory.

We derive quantitative relation between logical depth and
computational depth and unify the different {\em depth} notions by
relating them to A. Kolmogorov and L. Levin's fruitful notion of
randomness deficiency. Subsequently, we revisit the computational
depth of infinite strings, introducing the notion of {\it super deep}
sequences and relate it with other approaches.
\end{abstract}



\section{Introduction}
The information contained in an individual finite object (a finite
binary string) can be measured by its Kolmogorov complexity---the
length of the shortest binary program that computes the object. Such a
shortest program contains no redundancy: every bit is information; but
is it meaningful information? If we flip a fair coin to obtain a
finite binary string, then with overwhelming probability that string
constitutes its own shortest description. However, with overwhelming
probability also, all the bits in the string are apparently meaningless information, just random noise. 

The opposite of randomness is regularity; and the effective
regularities in an object can be used to compress it and cause it to
have lower Kolmogorov complexity. Regular objects contain laws that
govern their existence and have meaning. This meaning may be instantly 
clear, but it is also possible that this meaning becomes intelligible
only as the result of a long computation. For example, let the object in
question be a book on number theory. The book will list a number of
difficult theorems. However, it has very low Kolmogorov complexity
since all theorems are derivable from the initial few definitions. Our
estimate of the difficulty of the book is based on the fact that it
takes a long time to reproduce the book from part of the information
in it. We can transmit all the information in the book by just
transmitting the theorems. The receiver will have to spend a long time
reconstructing the proofs and the full book. On the other hand, we can
send all of the book. Now the receiver has all the useful information
without literally, and does not have to spend time to extract information. 
Hence, there is a tradeoff: in both
cases we send the same information in terms of Kolmogorov complexity,
but in the former case it takes a long time to reconstruct it from a
short message, and in the latter case it takes a short time to
reconstruct it from a long message. The existence of such book is
itself evidence of some long evolution preceding it. The computational effort
to transform the information into `usable' information is called `depth'.

We also use a central notion in Kolmogorov complexity: 
that of `randomness deficiency'.
The randomness deficiency of an object in a particular distribution quantifies the
`typicality' or `randomness' of that object for that distribution. 
A randomness deficiency of 0 tells us that the object is typical (we believe that the object is randomly drawn from the distribution). A high randomness deficiency
tells us that the object is atypical and not likely to be randomly drawn.
Finally, we consider the information in one object about another one
and vice versa, and since these are approximately equal we call it `mutual information.'

{\bf Results:} For finite strings, we derive 
quantitative relations between the different notions of
depth: logical depth and computational depth (Section 3). In Section 4
we prove that these two notions of depth are instances of a more general measure, namely, Levin's
randomness deficiency, i.e., computational depth is the randomness
deficiency with respect to the time bounded universal semimeasure and
 logical depth is the least time for wich the randomness deficiency with respect
to the time bounded apriori probability is upper bounded by the significance level.

Next, we study the information contained on
infinite sequences. Applying the randomness deficiency 
with respect to ${\bf M}\otimes {\bf M}$, where ${\bf M}$ 
is the universal lower semicomputable semimeasure over $\{0,1\}^{\infty}$,
Levin~ \cite{levin74,Lev84} defined mutual information for infinite sequences. We observe that
despite the correctness of the definition, it does not fully achieve
the desired characterization of mutual information. For example, if $\alpha=\alpha^1\alpha^2...$ and
$\gamma=\gamma^1\gamma^2...$ are two Kolmogorov random sequences
and we construct the sequence $\beta=\alpha^1\gamma^1\alpha^2\gamma^2...$, then
$I(\alpha:\beta)=I(\beta:\alpha)=\infty$. However intuitively $\beta$ has more
information about $\alpha$ than the other way around since from $\beta$ we can
fully reconstruct $\alpha$ but from $\alpha$ we can only recover half of
$\beta$. In order to fulfil our intuition we propose some definitions of
normalized mutual information for infinite sequences. We relate this notion with the constructive Hausdorff dimension, using
the result proved by Mayordomo in \cite{Mayordomo}. Namely, we show
that the normalized mutual information of $\alpha$ with respect to
$\beta$ is at least the ratio of the constructive Hausdorff dimensions
of $\alpha$ and $\beta$ up to an additive factor that measures the difficulty to recover the
initial segments of $\alpha$ from the initial segments of the same size of
$\beta$.
This connection motivates the definition of dimensional mutual
information for infinite sequences. This measure, contrarly to the normalized
mutual information, is symmetric and it is at most the minimum between
the  normalized mutual information of $\alpha$ with respect to $\beta$
and vice versa.

In the last section we revisit the notion of depth for infinite
sequences, proposing a new depth measure called dimensional depth. As the name
suggests, this measure is related to the constructive Hausdorff dimension. We prove that dimensional depth is at most the difference
between time bounded and resource unbounded versions of constructive Hausdorff
dimension and finally we fully characterize super deepness 
using our proposed measures in a similar way as done in
\cite{lutz}.

{\bf Previous work:}
Bennett~\cite{Bennett88} introduced the notion of logical depth of an
object as the amount of time required for an algorithm to derive the
object from a shorter description. 

Antunes {\it et al.}~\cite{afmv06} consider logical depth as one instantiation of a more general theme, computational depth, and propose several other variants based on the difference between a resource bound Kolmogorov complexity measure and the unbounded Kolmogorov complexity. 

For infinite sequences, Bennett identified the classes of {\it weakly}
and {\it strongly deep} sequences, and showed that the halting problem
is strongly deep. Intuitively a sequence is strongly deep if no
computable time bound is enough to compress infinitely many of its
prefixes to within a constant number of bits of its smallest
representation. An interpretation of strongly deep objects
is given in~\cite{ll99}; a strongly deep sequence is analogous to a
great work of literature for which no number of readings suffices to
exhaust its value. Subsequently Judes, Lathrop, and Lutz~\cite{lutz}
extended Bennett's work defining the classes of {\it weakly useful}
sequences. The computational usefulness of a sequence can be measured as the class of computational problems that can be
solved efficiently, given access to that sequence. More
formally, for infinite sequences, a sequence is weakly useful if every element of a non-negligible set of decidable
sequences is reducible to it in recursively bounded time. Lathrop, and
Lutz~\cite{lutz} proved that every weakly useful sequence is strongly
deep in the sense of Bennett. Later, Fenner {\it et al.}~\cite{flmr05} proved
that there exist sequences that are weakly useful but not strongly
useful. Lathrop and Lutz~\cite{ll99} introduced refinements (named
{\it recursive weak depth} and {\it recursive strong depth}) of
Bennett's notion of weak and strong depth, and studied its fundamental
properties, showing that recursively weakly (resp. strongly) deep sequences form a proper subclass of the
class of weakly (resp. strongly) deep sequences, and also that every
weakly useful sequences is recursive strongly deep.

 Levin~\cite{levin74,Lev84} showed that the randomness deficiency of $x$ with respect to $\mu$ is the largest, within an additive constant, randomness $\mu$-test for $x$. So $\delta (x \mid \mu)$ is, in a sense, a universal characterization of ``non-randomness'', ``useful'' or ``meaningful'' information in a string $x$ with respect to a probability distribution $\mu$.\bigskip

%
%

\section{Preliminaries}
We briefly introduce some notions from Kolmogorov complexity, mainly the standardize notation. We refer to the textbook by Li and Vit\'anyi~\cite{LiVi} for more details. Let $U$ be a fixed universal Turing machine. For technical reasons we choose one with a separate read-only input tape, that is scanned from left-to-right without backing up, a separate work tape on which the computation takes place, and a separate output tape. Upon halting, the initial segment $p$ of the input that has been scanned is called the input ``program'' and the contents of the output tape is called the ``output''. By construction, the set of halting programs is prefix free. We call $U$ the reference universal prefix machine. In the rest of this paper we denote the $n$- length prefix of an infinite sequence $\alpha$ by $\alpha_n$ and the $i$th bit by $\alpha^i$.

\begin{definition}\label{def.initial}
\rm
(i) The (prefix) {\em Kolmogorov complexity} of a finite binary string $x$ is defined as 
\[ K(x)=
\min_{p}\{|p| : U(p)= x \},
\]
where $p$ is a program, and the {\em Universal a priori probability} of $x$ is 
\[Q_U(x) = \sum_{U(p)=x} 2^{-|p|}.
\]

(ii) A time-constructible function $t$ from natural numbers to natural numbers is a function with the property that $t(n)$ can be constructed from $n$ by a Turing machine in time of order $O(t(n))$. For every time-constructible $t$, the $t$-{\em time-bounded Kolmogorov complexity} of $x$ is defined as \[
K^{t}(x) =
\min_p \{|p| : U(p) = x \text{ in at most } t(|x|)  \text{  steps}\},
\] 
and the $t$-{\em time bounded Universal a priori probability} is defined as 
\[
Q^t_U(x)=\sum_{U^t(p)=x}2^{-|p|},
\]
and $U^t(p)=x$ means that $U$ computes $x$ in at most $t(|x|)$ steps and halts. 
\end{definition}

A different universal Turing machine may affect the program size $|p|$ by 
at most a constant additive term, and the running time $t$ by at most a
logarithmic multiplicative factor. The same will hold for all other
measures we will introduce.

Levin~\cite{levin74} showed that the Kolmogorov complexity of a string $x$ coincides up to an additive constant term with the logarithm of $1/Q_U(x)$.
This result is called the ``Coding Theorem'' since it shows that the
shortest upper semicomputable code is a Shannon-Fano code of the
greatest lower semicomputable probability mass function. In order to
state formally the Coding theorem we need the following theorem on the
existence of a universal lower semicomputable discrete semimeasure (Theorem 4.3.1 in \cite{LiVi}).

\begin{theorem}
There exists a universal lower semicomputable discrete semimeasure over $\{0,1\}^*$, 
denoted by ${\bf m}$. 
\end{theorem}

\begin{theorem}[Coding Theorem]\label{LiVi}
For every $x\in \{0,1\}^n$,
$$\ds{K(x)= -\log Q_U (x)=-\log {\bf m}(x)}$$
with equality up to an additive constant $c$.
\end{theorem}

Hence, if $x$ has high probability because it has many long
descriptions then it must have a short description too.\bigskip

We refer to \textit{mutual information} of two finite strings as $$I(x:y)=K(x)+K(y)-K(x,y).$$

Notice that the mutual information is symmetric, i.e., $I(x:y)=I(y:x)$.

\section{Depth}

Bennett~\cite{Bennett88} defines the $b$-significant logical depth of
an object $x$ as the time required by the reference universal Turing
machine to generate $x$ by a program that is no more than $b$ bits
longer than the shortest descriptions of $x$. Bennett talks about time
as the number of steps; without loss of generality we consider the
number of steps $t(|x|)$, where $t$ is a time-constructible function.

\begin{definition}[Logical Depth]\label{def.2}
\rm
The logical depth of a string $x$ at a significance level $b$ is
\[
\ldepth_b(x)=\min \left\{t(|x|): \frac{Q^t_U(x)}{Q_U(x)} \geq 2^{-b}\right\},
\]
where the minimum is taken over all time constructible $t$. 
\end{definition}

Given a significance level $b$, the logical depth of a string $x$ is the minimal running time $t(|x|)$, such that programs running in at most $t(|x|)$ steps account for approximately a $1/2^b$ fraction of $x$'s universal probability.
This is Bennett's Tentative Definition 0.3 in \cite{Bennett88} p. 240.

In fact, with some probability we
can derive the string by simply flipping a coin. But for long strings
this probability is exceedingly small. If the string has a short description then we can flip that description  with higher probability. Bennett's proposal tries to
express the tradeoff between the probability of flipping a short
program and the shortest computation time from program to object. 

Antunes {\it et al.}~\cite{afmv06} developed the notion of computational depth in order to capture the tradeoff between the amount of help bits required and the reduced computation time to compute a string. 
The concept is simple: they consider the difference of two versions of Kolmogorov complexity measures.

\begin{definition}[Basic Computational Depth]
\rm
Let $t$ be a time constructible function. For any finite binary string $x$
we define $$\depth^t(x)=K^{t}(x)-K(x).$$
\end{definition}

In Definition 1 of
\cite{Bennett88} p. 241 we find 
\begin{definition}\label{def.1}
\rm
A string $x$ is $(t(|x|),b)$-deep iff $t(|x|)$ is the least number of steps to compute $x$ from a program of length at most $K(x)+b$. 
\end{definition}
Then, it is straightforward that $\depth^t(x)=K^t(x)-K(x)$ iff $x$ is $(t(|x|), K^{t}(x)-K(x))$-deep. Bennett remarks, \cite{Bennett88} p. 241,
``The difference between [Definitions~\ref{def.1} and \ref{def.2}] 
is rather subtle philosophically and not very great quantitatively.''
This is followed by \cite{Bennett88} Lemma 5 on p. 241 which is an informal
version of \cite{LiVi} Theorem 7.7.1.
The proof of Item (ii) below uses an idea in 
the proof of the latter theorem. 
\begin{definition}\label{def.Kt}
\rm
Let $t$ be a recursive function.
Define $K(t)$ as the (prefix) Kolmogorov complexity of 
$t$ by $K(t)=\min_i \{i: T_i$ computes $t(\cdot)\}$, where $T_1, T_2, \ldots$ is the standard enumeration of all Turing machines.
\end{definition}

\begin{theorem}\label{theo.1}
Let $t$ be a time-constructible function (hence it is recursive and
$K(t)$ is defined in Definition~\ref{def.Kt}).

(i) If $b$ is the minimum value such that $\ldepth_b(x) = t(|x|)$, then  $\depth^t(x) \geq b+O(1)$.

(ii) If $\depth^{t}(x)=b$, then $\ldepth_{b+\min\{K(b),K(t)\}+O(1)}(x) \geq t(|x|)$.
\end{theorem}
\begin{proof}
(i) 
Assume, $\ldepth_{b}(x) = t(|x|)$. So 
\[
\frac{Q^t_U(x)}{Q_U(x)} \geq 2^{-b},
\]
with $t(|x|)$ least. Assume furthermore that  $b$ is the least integer
so that the inequality holds for this $t(|x|)$.
We also have 
\[
\frac{Q^t_U(x)}{Q_U(x)} \geq \frac{2^{-K^{t}(x)}}{Q_U(x)}
= 2^{-(K^t(x)-K(x)-O(1))} = 2^{-b - \Delta},
\]
where 
$b + \Delta =  K^t(x)-K(x)-O(1)$. The first inequality holds since
the sum $Q^t_U(x)$ comprises a term $2^{-K^{t}(x)}$
based on a shortest program of length $K^{t}(x)$
computing $x$ in at most $t(|x|)$ steps. Since $b$ is the least integer, it follows that $\Delta \geq 0$. Since $\depth^{t}(x) = K^t(x)-K(x)$, we find that
$\depth^{t}(x) \geq  b+O(1)$.\bigskip

(ii) 
Assume that  $\depth^{t}(x)=b$, that is,  $x$ is
$(t(|x|), b)$-deep.
We can enumerate the set $S$ of all programs computing $x$ in time at most $t(|x|)$
by simulating all programs of length $l \leq |x|+2 \log |x|$ for $t(|x|)$
steps. Hence, the shortest such program $q$ enumerating $S$
has length $|q| \leq K(x,t)+O(1)$. But we achieve the same effect
if, given $x$ and $b$ we enumerate all programs of length $l$ as above
in order of increasing running time and stop when the accumulated
algorithmic probability exceeds $2^{-K(x)+b}$. The running time of the last
program is $t(|x|)$. (This shows that $K(t,x) \leq K(b,x)+O(1)$,
not $K(t) \leq K(b)+O(1)$). The shortest program $r$ doing this
has length $|r|\leq K(x,b)+O(1)$. Hence, 
$K(S) \leq \min \{K(x,t),K(x,b)\}+O(1)$.
By definition,
$Q^t_U(x) = \sum_{p \in S} 2^{-|p|}$.
Assume, by way of contradiction, that
\[
 \frac{Q^t_U(x)}{Q_U(x)} < 2^{-b - \min\{K(b),K(t)\}-O(1)}
\]
Since $Q_U(x) = 2^{-K(x)-O(1)}$, we 
have 
\[
Q^t_U(x) < 2^{-K(x)-b-\min \{K(b),K(t)\}-O(1)}
\]
Denote $m=K(x)+b+\min\{K(b),K(t)\}+O(1)$.
Therefore,
$\sum_{p \in S} 2^{-|p|} < 2^{-m}$.
Now every string in $S$ can be effectively compressed by
at least $m-K(S)-O(1)$ bits.
Namely, 
\[
\sum_{p \in S} 2^{-|p| +m } < 1
\]
The latter inequality is a Kraft inequality,
and hence the elements of $S$ can be coded by a prefix code 
with the code word length for $p$ at most $|p|-m$. 
In order to make this coding effective, we use a program of length $K(S)$ to enumerate exactly the strings of $S$. This takes an additional $K(S) + O(1)$ bits in the code for each $p \in S$. This way, each $p \in S$ is effectively compressed by $m-K(S)-O(1)$ bits. Therefore, each $ p \in S$ can be compressed by at least $K(x)+b+\min\{K(b),K(t)\}-\min\{K(x,t),K(x,b)\}$ bits, up to an additive constant we can set freely, and hence by more than $b$ bits which is a contradiction.
Hence, 
\[
\frac{Q^t_U(x)}{Q_U(x)} \geq 2^{-b - \min\{K(t),K(b)\}-O(1)}
\]
which proves (ii).
\end{proof}

\section{A Unifying Approach}

Logical depth and computational depth are all instances of a more general measure, namely the randomness deficiency of a string $x$ with respect to a probability distribution, Levin \cite{levin74,Lev84}. In the rest of this paper, with some abuse of notation (see \cite{LiVi}), a function $\mu : \{0,1\}^* \rightarrow \mathbb R$ defines a {\em probability measure}, or {\em measure} for short, if
\begin{eqnarray*}
&& \mu  (  \epsilon )   =  1, \\
&& \mu (x)   =   \sum_{{a}  \in  \{0,1\}} \mu ( x a ).
\end{eqnarray*}

 \begin{definition}
\rm
 Let $\mu$ be a computable measure. The value 
\[
\delta (x \mid \mu)= \left\lfloor \log \frac{Q_U(x)}
     {\mu(x)} \right\rfloor 
\]
is the {\em randomness deficiency}\footnote{$\lfloor r \rfloor$ denotes the integer part of $r$  and $\lceil \alpha \rceil$ denotes the smallest integer bigger than $\alpha$.} of $x$ with respect to $\mu$. Here $Q_U$ is the universal a priori probability of Definition \ref{def.initial}. 
 \end{definition}

Note that $Q_U(x)$ is of exact order of magnitude of $2^{-K(x)}$ by the Coding Theorem~\ref{LiVi}, i.e., up to multiplicative terms $Q_U(x)$ and $2^{-K(x)}$ are equal. (In the literature, see for example \cite{LiVi}, ${\bf m}(x)= 2^{-K(x)}$ is used instead of $Q_U(x)$, and it is straightforward that this is equivalent up to a multiplicative independent constant by the Coding Theorem.)\bigskip

We now observe that logical depth and computational depth of a string $x$ equals the randomness deficiency of $x$ with respect to the measures $Q^t(x)=\sum_{U^t(p)=x}2^{-|p|}$ and $2^{-K^t(x)}$ respectively. The proofs follow directly from the definitions.

\begin{lemma}
Let $x$ be a finite binary string and let $t$ be a time-constructible
function.

(i) $ \ldepth_b(x)=\min \{t:\delta (x \mid Q^t) \leq b \}$.
 
(ii) $\depth^{t}(x) = \delta (x \mid {\bf m}^t)$ where ${\bf m}^t(z)=2^{-K^t(z)}$.
 \end{lemma}


\section{On the information of infinite strings}

Based on the unification of depth concepts for finite strings, in this
section we extend those ideas for infinite sequences. In order to motivate our approach we start by introducing Levin's notion of randomness deficiency for infinite sequences.
Let ${\bf M}$ be the universal lower semicomputable 
(continous) semimeasure over $\{0,1\}^{\infty}$
as defined, and proved to exist, by \cite{Lev84} (see also \cite{LiVi}).
If $\alpha \in \{0,1\}^{\infty}$, then with $\alpha = \alpha^1 \alpha^2 \ldots$ with
$\alpha^i \in \{0,1\}$, we write $\alpha_n = \alpha^1 \alpha^2 \ldots \alpha^n$. 
Finally,  we write `${\bf M}(x)$' and `$\mu(x)$' as notational shorthand
for `${\bf M}(\Gamma_x)$' and `$\mu(\Gamma_x)$', with $x \in \{0,1\}^*$ 
and $\Gamma_x$ is the {\em cylinder} 
$\{\omega: \omega \in \{x\}\{0,1\}^{\infty} \}$.
Strictly speaking, ${\bf M}(x)$ is not over $\{0,1\}^{\infty}$ but over
$\{0,1\}^{\infty} \bigcup  \{0,1\}^*$, see also \cite{LiVi},
and ${\bf M}(x)$ is the probability concentrated on the set of finite and infinite sequences 
starting with $x$.

\begin{definition}[Levin]
\rm
The value $\displaystyle{D(\alpha \mid \mu)=\left \lfloor \log \left (
      \sup_n \frac{{\bf M}(\alpha_n)}{\mu(\alpha_n)} \right )\right
  \rfloor}$ is called the randomness deficiency of $\alpha$ with
respect to the semimeasure $\mu$. Here ${\bf M}(\alpha_n)$ is the
probability density function of ${\bf M}(\alpha_n)$.
\end{definition}

Let $\alpha$ and $\beta$ be two sequences and ${\bf M} \otimes {\bf M}$ be defined by ${\bf M} \otimes {\bf M}(\alpha,\beta)={\bf M}(\alpha){\bf M}(\beta)$.

\begin{definition}[Levin]\label{def_l}
\rm
The value $I(\alpha:\beta)=D((\alpha,\beta) \mid {\bf M}\otimes {\bf M})$ is called the amount of information in $\alpha$ about $\beta$ or the deficiency of their independence.
\end{definition}

This definition is equivalent to the mutual information $I(\alpha:\beta)=\sup_{n} I(\alpha_n:\beta_n)$.\bigskip

\begin{example}\label{ex}
\rm
Let $\alpha$ and $\gamma$ be two random infinite and independent strings (in the sense that their prefixes are independent). Consider the following sequence

\[
\beta=\alpha^1\gamma^1\alpha^2\gamma^2\ldots
\]
By Definition~\ref{def_l} we have
$$\begin{array}{lll}
I(\alpha:\beta)&=&\ds{\sup_n I(\alpha_n:\beta_n)}\\
&=&\ds{\sup_n(K(\alpha_n)+K(\beta_n)-K(\alpha_n,\beta_n))}\\
&\geq& \ds{\sup_n\left(n+n-\left(n+\frac{n}{2}\right)\right)=\infty.}
\end{array}$$ 
As $I(\beta:\alpha)=I(\alpha:\beta)$ then $I(\beta:\alpha)=\infty$.\bigskip

However, intuitively $\beta$ contains more information about $\alpha$ than the other way around, since from the sequence $\beta$ we can totally reconstruct $\alpha$ but from $\alpha$ we can only recover half of $\beta$, namely, the bits with odd indexes.
\end{example}

This seems to be a lacuna in Definition~\ref{def_l}. The definition says more when the information is finite but that is precisely when we do not need an accurate result. Notice that if the sequences are finite we can argue that they are independent. In the infinite case, one should be able to classify the cases where the mutual information is infinite. Two infinite sequences may have infinite mutual information and yet infinite information may be still lacking to reconstruct one of them out of the other one. In the previous example $\alpha$ fails to provide all the information of $\beta$ related to $\gamma$, which has infinite information. In this section we will present two approaches to reformulate the definition of ``mutual information'' in order to fulfill our intuition. In order to have a proportion of information as the prefixes grow we need to do some normalization in the process.

\subsection{The Mutual Information Point of View}

We are looking for a normalized mutual information measure $I_m$ that applied to Example~\ref{ex} gives	\[
   I_m(\alpha:\alpha)=1
\]
\[
I_m(\alpha:\beta)=1/2;
\]
\[
I_m(\beta:\alpha)=1
\]
\[
I_m(\beta:\beta)=1
\]
Contrarily to Levin's definition of mutual information for infinite
sequences, and accordingly to our intuition, the above conditions
imply that the normalized version must be non-symmetric.

\begin{definition}[First attempt] 
\rm
Given two infinite sequences $\alpha$ and $\beta$ the normalized
mutual information that $\beta$ has about $\alpha$ is defined as
  \[
  I_m(\beta:\alpha)=\lim_{n\rightarrow\infty}\lim_{m\rightarrow\infty}\frac{I(\beta_m:\alpha_n)}{I(\alpha_n:\alpha_n)}
  \]
\end{definition}

The major drawback of this definition is the fact that the limit does not always exist.\footnote{Notice that there are sequences $\alpha$ for which $\displaystyle{\lim_n\frac{n}{K(\alpha_n)}}$ does not exist.} However, it does exist for the Example~\ref{ex} with the desired properties. Furthermore, we obtain for the same $\alpha$ and $\beta$
\[
I_m(\alpha:\alpha)=1;
\]
\[
I_m(\beta:\beta)=1;
\]
\[
I_m(\alpha:\beta)=\lim_{n\rightarrow\infty}\lim_{m\rightarrow\infty}\frac{m+n-(m+n-n/2)}{n}=\frac{1}{2};\]
\[I_m(\beta:\alpha)=\lim_{n\rightarrow\infty}\lim_{m\rightarrow\infty}\frac{m+n-m}{n}=1;
\]

\begin{definition}[Normalized mutual information for infinite sequences]
\rm
Given two \; infinite sequences $\alpha$ and $\beta$ we define the lower normalized mutual information that $\beta$ has about $\alpha$ as
\[
I_{m*}(\beta:\alpha)=\liminf_{n\rightarrow\infty}\lim_{m\rightarrow\infty}\frac{I(\beta_m:\alpha_n)}{I(\alpha_n:\alpha_n)}
\]
and the upper normalized mutual information that $\beta$ has about $\alpha$ as
\[
I_m^*(\beta:\alpha)=\limsup_{n\rightarrow\infty}\lim_{m\rightarrow\infty}\frac{I(\beta_m:\alpha_n)}{I(\alpha_n:\alpha_n)}
\]
\end{definition}

Notice that these definitions also fulfill the requirements presented
in the beginning of this section with respect to Example \ref{ex}.

We now can define independence with respect to normalized mutual information:

\begin{definition}
\rm
Two sequences, $\alpha$ and $\beta$, are independent if $I_m^*(\alpha:\beta)=I_m^*(\beta:\alpha)=0$. 
\end{definition}

In \cite{Lutz00,Lutz02}, the author developed a constructive version
of Hausdorff dimension. That dimension assigns to every binary
sequence $\alpha$ a real number $\dim(\alpha)$ in the interval
$[0,1]$. Lutz claims that the dimension of a sequence is a measure of
its information density. The idea is to differentiate sequences by
non-randomness degrees, namely by their dimension. Our approach is
precisely to introduce a measure of density of information that one
sequence has about the other, in the total amount of the other's
information. So we differentiate non-independent sequences, by their
normalized mutual information.

Mayordomo \cite{Mayordomo} redefined constructive Hausdorff dimension in terms of Kolmogorov complexity.

\begin{theorem}[Mayordomo]
For every sequence $\alpha$, \[\dim(\alpha)=\liminf_{n\rightarrow\infty}\frac{K(\alpha_n)}{n}\]
\end{theorem}

So, now the connection between constructive dimension and normalized
information measure introduced here is clear. It is only natural to
accomplish results about the Hausdorff constructive dimension of a
sequence, knowing the dimension of another, and their normalized
information. 

\begin{lemma}
Let $\alpha$ and $\beta$ be two infinite sequences. Then
	\[I_m^*(\alpha:\beta)\cdot\dim(\beta)\geq \dim(\alpha) +\liminf_{n\to\infty}-\frac{K(\alpha_n|\beta_n)}{n}
\]
\end{lemma}
\begin{proof}
\[
\begin{array}{lll}
I_m^*(\alpha:\beta)\cdot\dim(\beta)&=&\ds{\limsup_n\lim_m\frac{I(\alpha_m:\beta_n)}{I(\beta_n:\beta_n)}\cdot\liminf_{n}\frac{K(\beta_n)}{n}}\\
&\geq&\ds{\liminf_n\liminf_m\frac{I(\alpha_m:\beta_n)}{n}}\\
&\geq&\ds{\liminf_n\liminf_m\frac{I(\alpha_m:\beta_n)}{m}}\\
&\geq&\ds{\liminf_n\liminf_m\frac{K(\alpha_m)-K(\alpha_m|\beta_m)}{m}}\\
&\geq&\ds{\liminf_m\frac{K(\alpha_m)}{m}+\liminf_m\frac{-K(\alpha_m|\beta_m)}{m}}\\
&=&\ds{\dim(\alpha)+\liminf_m-\frac{K(\alpha_m|\beta_m)}{m}}
\end{array}
\]
\end{proof}

Note that, in the previous lemma the (unexpected) aditive term
$\displaystyle{\liminf_m-\frac{K(\alpha_m|\beta_m)}{m}}$ is necessary
to expresses the hardness of recover $\alpha$ given $\beta$.

We present now the time bounded version of $\dim(\alpha)$. This definition will be important later on this paper.

\begin{definition}
\rm
Let $t$ be a time-constructible function.
The $t$-bounded dimension of an infinite sequence $\alpha$ is defined as
\[
\dim^t(\alpha)=\liminf_{n\to\infty}\frac{K^t(\alpha_n)}{n}
\]
\end{definition}

\subsection{The Hausdorff constructive dimension point of view}
In this subsection we define a version of mutual information between two
sequences based on Hausdorff constructive dimension and establish a connection to it.

\begin{definition}
\rm
The dimensional mutual information of the sequences $\alpha$ and
$\beta$ is defined as 
	\[I_{dim}(\alpha:\beta)=\dim(\alpha) + \dim(\beta)
													-2\dim \left<\alpha,\beta \right>
\]
\end{definition}
This measure of mutual information is symmetric. The definition considers twice $\dim \left <\alpha,\beta \right >$ because when encoding the prefixes $\alpha_n$ and $\beta_n$ the result is a $2n$-length string. 
Notice that, 
\begin{eqnarray*}
I_{dim}(\alpha: \beta) 
& = & \dim(\alpha) + \dim(\beta) - 2\dim \left <\alpha,\beta\right >\\ 
& = & \liminf_{n\rightarrow \infty} \frac{K(\alpha_{n/2})}{n/2} +
\liminf_{n\rightarrow
  \infty}\frac{K(\beta_{n/2})}{n/2}-2\liminf_{n\rightarrow
  \infty}\frac{K(\left <\alpha,\beta\right>_n)}{n}\\
& \leq & \liminf_{n\rightarrow \infty}\frac{K(\alpha_{n/2})+K(\beta_{n/2})-K(\alpha_{n/2},\beta_{n/2})}{n/2}\\
& = & \liminf_{n\rightarrow \infty}\frac{I(\alpha_n:\beta_n)}{n}\\
&\leq &\liminf_{n\rightarrow \infty}\frac{I(\alpha_n:\beta_n)}{K(\beta_n)}\\
&\leq
 & \liminf_{n\rightarrow \infty}\lim_{m\rightarrow \infty}\frac{I(\alpha_m:\beta_n)}{K(\beta_n)}\\
&=&I_{m*}(\alpha:\beta)
\end{eqnarray*}

The third inequality is true due to the following fact:
\[I(\beta_n:\alpha_m)=K(\beta_n)-K(\beta_n|\alpha_m)\geq K(\beta_n)-K(\beta_n|\alpha_n)=I(\beta_n:\alpha_n).\]

By the symmetry of the definition we also have that $I_{dim}(\alpha: \beta)\leq I_{m*}(\beta:\alpha)$. These two facts prove the following lemma:

\begin{lemma}
Let $\alpha$ and $\beta$ be two sequences. Then $$I_{dim}(\alpha: \beta)\leq \min(I_{m*}(\alpha:\beta),I_{m*}(\beta:\alpha))$$
\end{lemma}


One can easily modify the definitions introduced in this section by considering the limits when $n$ goes to the length of the string, or the maximum length  of the strings being considered. One should also notice that when $x$ and $y$ are finite strings and $K(y)\geq K(x)$, $I_{m*}(x:y)$ is $1- d(x,y)$, where $d(x,y)$ is the normalized information distance studied in \cite{Li03}.

\section{Depth of infinite strings}

In this section we revisit depth for infinite sequences. We introduce a new depth
measure, prove that it is closely related with constructive
Hausdorff dimension and use it to characterize super deepness. To
motivate our definitions we recall the definitions of the classes of
weakly (vs. strongly deep) sequences and weakly useful (vs. strongly useful)
sequences.

\begin{definition}[\cite{Bennett88}]
\rm
An infinite binary sequence $\alpha$ is defined as
\begin{itemize}
\item weakly deep if it is not computable in recursively
bounded time from any algorithmically random infinite sequence.
\item strongly deep if at every significance level
$b$, and for every recursive function $t$, all but finitely many initial segments
$\alpha_n$ have logical depth exceeding $t(n)$.
\end{itemize}
\end{definition}

\begin{definition}[\cite{flmr05}]
\rm
An infinite binary sequence $\alpha$ is defined as
\begin{itemize}
\item weakly useful if there is a computable time bound within which all the sequences in a non-measure $0$ subset of the set of decidable sequences are Turing reducible to $\alpha$.
  \item strongly useful if there is a computable time bound within which every decidable sequence is Turing reducible to $\alpha$.
\end{itemize}
\end{definition}

The relation between logical depth and usefulness was studied by
Juedes, Lathrop and Lutz~\cite{lutz} who defined the conditions for
weak and strong usefulness and showed that every weakly useful
sequence is strongly deep. This result generalizes Bennett's
remark that the diagonal halting problem is strobgly deep, strengthening the relation between depth and usefulness. Latter Fenner {\it et al.}~\cite{flmr05} proved the existence of sequences that are weakly useful but not strongly useful.

The Hausdorff constructive dimension has a close connection with the information theories for infinite strings studied before, see for example \cite{flmr05}, \cite{Lutz00}, \cite{Lutz02} and \cite{Mayordomo}. Therefore, in this section we define the dimensional computational depth of a sequence in order to study the nonrandom information on a infinite sequence.

\begin{definition}
\rm
The dimensional depth of a sequence $\alpha$ is defined as
	\[\depth^t_{\dim}(\alpha)=\liminf_{n\to\infty}\frac{\delta(\alpha_n  \mid  2^{-K^t(\alpha_n)})}{n}.
\]
\end{definition}


\begin{lemma}
	\[depth^t_{\dim}(\alpha)\leq \dim^t(\alpha)-\dim(\alpha)
\]

\end{lemma}
\begin{proof}
\[\begin{array}{lll}
\depth^t_{\dim}(\alpha)&=&\ds{\liminf_{n\to\infty}\frac{\delta(\alpha_n  \mid 
 2^{-K^t(\alpha_n)})}{n}}\\ &=& \ds{\liminf_{n\to\infty}\frac{K^t(\alpha_n)-K(\alpha_n)}{n}}\\ &\leq& \ds{\dim^t(\alpha) - \dim(\alpha)}.\end{array}
\]
The last inequality holds since the sequence of values $K(\alpha_n)/n$ is non negative and then $\ds{\liminf_n -K(\alpha_n)/n\leq -\liminf_n K(\alpha_n)/n}$.
\end{proof}

%
%


%




Now, in the definition of strongly deep sequences, instead of considering a fixed significance level $s$ we consider a significance level function $s:\mathbb{N}\rightarrow\mathbb{N}$ . Naturally, we want $s(n)$ to grow very slowly so we assume for example that $s=o(n)$. With this replacement we obtain a tighter definition as deepness decreases with the increase of the significance level.

\begin{definition}
\rm
A sequence is called super deep if for every significance level
function\linebreak $s:\mathbb{N}\rightarrow\mathbb{N}$, such that $s=o(n)$, and
for every recursive function $t:\mathbb{N}\rightarrow\mathbb{N}$, all
but finitely many initial segments $\alpha_n$ have logical depth
exceeding $t(n)$.
\end{definition}

We have already characterized super deep sequences using their dimensional depth in Theorem \ref{theo.1}. In fact we have

$$\ldepth_b(x) = t(|x|)\textnormal{, with $b$ minimal}\Rightarrow \depth^t(x) \geq b+O(1)$$

\begin{theorem}\label{sd}
A sequence $\alpha$ is super deep if and only if $\depth_{\dim} ^t(\alpha)>0$ for all recursive time bound $t$.
\end{theorem}

\begin{proof}
Let $\alpha$ be a super deep sequence. Then for every significance level function $s$, such that $s=o(n)$ and every recursive function $t$ we have that for almost all $n$, $\ldepth_{s(n)}(\alpha_n)>t(n)$. Then
\[
\depth^{t(n)}(\alpha_n)>s(n).
\]

Now if for some time bound $g$, $\depth^g_{\dim} (\alpha)=0$
then there exists a bound $S$, such that $S=o(n)$,
and, infinitely often

$$\depth^{g(n)}(\alpha_n)<S(n).$$
This is absurd and therefore for all recursive time bound
$t$, $\depth^t_{\dim}(\alpha)>0$.

Conversely if $\depth^t_{\dim}(\alpha)>0$ then there is some
$\epsilon > 0$ such that for almost all $n$, $\depth^{t(n)}_{\dim}(\alpha_n)>\epsilon n$. This implies that
$$\ldepth_{s(n)}(\alpha_n)>\ldepth_{\epsilon n}(\alpha_n)>t(n)$$ for all
significance function $s=o(n)$ and almost all $n$. So $\alpha$ is
super deep.
\end{proof}
\\

In the next theorem we express other equivalent ways to define super deepness. 

\begin{theorem}\label{equiv}
For every sequence $\alpha$ the following conditions are equivalent.
\begin{enumerate}
	\item $\alpha$ is super deep;
	\item For every recursive time bound $t:\mathbb{N}\rightarrow \mathbb{N}$ and every significance function $g=o(n)$, $\depth^t(\alpha_n)>g(n)$ for all except finitely many $n$;
	\item For every recursive time bound $t:\mathbb{N}\rightarrow \mathbb{N}$ and every significance function $g=o(n)$, $Q(\alpha_n)\geq 2^{g(n)}Q^t(\alpha_n)$ for all except finitely many $n$;
\end{enumerate}
\end{theorem}

\begin{proof}[Sketch] The equivalence $(1\Leftrightarrow 2)$ was proved in Theorem \ref{sd}. To show that $(2\Leftrightarrow 3)$ consider the following sets:
\[D^t_g=\{\alpha\in\{0,1\}^{\infty}:\depth^t(\alpha_n)\geq g(n)\textnormal{ a.e.}\}\]
\[\tilde{D}^t_g=\{\alpha\in\{0,1\}^{\infty}:Q(\alpha_n)\geq2^{g(n)}Q^t(\alpha_n)\textnormal{ a.e.}\}\]
The proof nows is an immediate consequence of the following lemma:

\begin{lemma}[Lemma 3.5 in \cite{lutz}]
If $t$ is a recursive time bound then there exists constants $c_1$ and $c_2$ and a recursive time bound $t_1$ such that $D^{t_1}_{g+c_1}\subset \tilde{D}^t_g$ and $\tilde{D}^t_{g+c_2}\subset D^t_g$.
\end{lemma}

\end{proof}

Following the ideas in \cite{lutz} to prove that every weakly useful sequence is strongly deep we can prove that every weakly useful sequence is super deep. 

\begin{theorem}\label{super}
Every weakly useful sequence is super deep.
\end{theorem}

For the proof of this result we need the following lemmas:

\begin{lemma}[Lemma 5.5 in \cite{lutz}]\label{lemma55}
Let $s:\mathbb{N}\rightarrow\mathbb{N}$ be strictly increasing and time-constructible with the constant $c_s$ as witness. For each s-time-bounded Turing machine M, there is a constant $c_M$ that satisfies the following. Given non-decreasing functions $t, g:\mathbb{N}\rightarrow\mathbb{N}$ we define $s^*, \tau, \hat{t}, \hat{g}:\mathbb{N}\rightarrow\mathbb{N}$ by
$$s^*(n)=2^{s(\left\lceil \log n\right\rceil)+1},$$
$$\tau(n)=t(s^*(n+1)+4s^*(n+1)+2(n+1)c_s s(|w|)+2ns^*(n+1)s(|w|)),$$
$$\hat{t}=c_M(1+\tau(n)\left\lceil \log\tau (n)\right\rceil),$$
$$\hat{g}=g(s^*(n+1))+c_M,$$
where $w$ is the binary representation of $n$. For all sequences $\alpha$, $\beta$, if $\beta$ is Turing reducible to $\alpha$ in time $s$ by $M$ and $\beta\in D^{\hat{t}}_{\hat{g}}$ then $\alpha\in D^t_g$.
\end{lemma}

\begin{lemma}[Corollary 5.9 in \cite{lutz}]\label{lemma59}
For every recursive function $t:\mathbb{N}\rightarrow\mathbb{N}$ and every $0<\gamma<1$, the set $D_{\gamma n}^t$ has measure $1$ in the set of recursive sequences.
\end{lemma}

\begin{proof}[of Theorem \ref{super}]
Let $\alpha$ by a weakly useful sequence. To prove that $\alpha$ is super deep we show that for every recursive time bound $t$ and every any significance level $g=o(n)$, $\alpha\in D^t_g$, where $D^t_g$ is the set defined in proof of Theorem \ref{equiv}.

Since $\alpha$ is weakly useful then there exists a recursive time bound $s$ (that without lose of generality we can assume increasing) such that the set $DTIME^{\alpha}(s)$ of all sequences that are Turing reducible to $\alpha$ has positive measure in the set of recursive sequences. Using Lemma \ref{lemma55}, to conclude that $\alpha\in D^t_g$ all that is necessary is to prove that there exists $\beta\in D^{\hat t}_{\hat g}\cap DTIME^{\alpha}(s)$, where $\hat t$ and $\hat g$ are described in same lemma.

Fix $\gamma\in]0,1[$ and consider $\tilde t(n)=n(1+\gamma(n)\lceil\log
n\rceil)$ where $\gamma$ is obtained from $t$ and $s$ as in Lemma
\ref{lemma55}. Since $\tilde t$ is recursive, by Lemma \ref{lemma59},
$D^{\tilde t}_{\gamma n}$ has measure $1$ in the set of all recursive
sequences. Thus $D^{\tilde t}_{\gamma n}\cap DTIME^{\alpha}(s)$ has
measure $1$ and in particular is non empty. As $\tilde t(n)>\hat t(n)$
a.e. and $\gamma n>o(n)=g$ a.e. it follows, directly from the
definitions, that $D^{\tilde t}_{\gamma n}\subset D^{\hat t}_{\hat g}$
and then $D^{\tilde t}_{\gamma n}\not=\emptyset$, as we wanted to show.
\end{proof}

\begin{corollary}
The characteristic sequences of the halting problem and the diagonal halting problem are super deep.
\end{corollary}

\begin{proof} In \cite{Bennett88}, the author proved that the characteristic sequences of the halting problem and the diagonal halting problem are weakly useful. Then, it follows from Theorem \ref{super} that these two sequences are super deep.
\end{proof}

\subsection*{Acknowledgement}
We thank Harry Buhrman, Lance Fortnow, and Ming Li 
for comments and suggestions.

\bibliographystyle{alpha}

\begin{thebibliography}{}


\bibitem[AFMV06]{afmv06}
L. Antunes, L. Fortnow, D. van Melkebeek and N. Vinodchandran,
\newblock {\it ``Computational depth: concept and applications''},
\newblock in {\em Theor. Comput. Sci.}, volume 354 (3), pages: 391-404, 2006.

\bibitem[Ben88]{Bennett88}
C. Bennett,
\newblock{\it ``Logical depth and physical complexity''},
\newblock in {\em The Universal Turing Machine: A
  Half-Century Survey}, pages: 227-257, Oxford University Press, 1988.

\bibitem[FLMR05]{flmr05}
S. Fenner, J. Lutz, E Mayordomo and P. Reardon,
\newblock {\it ``Weakly useful sequences''},
\newblock in {\em Information and Computation} volume 197, pages: 41-54, 2005.


\bibitem[JLL94]{lutz}
D. Juedes, J. Lathrop and J. Lutz,
\newblock {\it ``Computational Depth and Reducibility''},
\newblock in {\em Theoret. Comput. Sci.}, volume 132, pages: 37-70, 1994.

\bibitem[LL99]{ll99} 
J. Lathrop and J. Lutz,
\newblock {\it ``Recursive computational depth''},
\newblock in {\em Information and Computation}, volume 153, pages: 139-172, 1999

\bibitem[Lev74]{levin74}
L. Levin,
\newblock {\it ``Laws of information conservation (nongrowth) and aspects of the foundation of probability theory''},
\newblock in {\em Probl. Inform. Transm.}, volume 10, pages: 206-210, 1974.

\bibitem[Lev84]{Lev84}
L. Levin,
\newblock {\it ``Randomness conservation inequalities: information and independence in mathematical theories''},
\newblock in {\em Information and Control}, volume 61, pages:15-37, 1984.

\bibitem[Li03]{Li03}
 M. Li, X. Chen, X. Li, B. Ma and P. Vit\'anyi,
\newblock {\it ``The similarity metric''},
\newblock {\em IEEE Trans. Inform. Th.},  50:12(2004), 3250--3264. 

\bibitem[LV97]{LiVi}
M. Li and P. Vit\'anyi,
\newblock {\it ``An introduction to Kolmogorov complexity and its
  applications''},
\newblock Springer, 2nd edition, 1997.

\bibitem[Lut00]{Lutz00}
J. Lutz,
\newblock {\it ``Dimension in complexity classes''},
\newblock {\em Proceedings of the 15th IEEE Conference of Computational Complexity}, IEEE Computer Society Press, 2000.

\bibitem[Lut02]{Lutz02}
J. Lutz,
\newblock {\it ``The dimensions of individual strings and sequences''},
\newblock {\em Technical Report cs.CC/0203017, ACM Computing Research Repository}, 2002.


\bibitem[May02]{Mayordomo}
E. Mayordromo,
\newblock {\it ``A Kolmogorov complexity characterization of constructive Hausdorff dimension''},
\newblock {\em Information Processing Letters}, volume 84, pages:1-3, 2002.


\end{thebibliography}

\end{document}